\documentclass[%
 aip,
 amsmath,amssymb,
 reprint,%
]{revtex4-1}

\usepackage{graphicx}% Include figure files
\usepackage{dcolumn}% Align table columns on decimal point
\usepackage{bm}% bold math
\usepackage[utf8]{inputenc}
\usepackage[T1]{fontenc}
\usepackage{mathptmx}
\usepackage{etoolbox}

\usepackage{amsmath}
\usepackage[colorlinks=true, linkcolor=blue, urlcolor=blue, citecolor=blue]{hyperref} % Clickable links
\usepackage{url}
\usepackage{subcaption}
\usepackage{orcidlink}

\usepackage{placeins}

\makeatletter
\def\@email#1#2{%
 \endgroup
 \patchcmd{\titleblock@produce}
  {\frontmatter@RRAPformat}
  {\frontmatter@RRAPformat{\produce@RRAP{*#1\href{mailto:#2}{#2}}}\frontmatter@RRAPformat}
  {}{}
}%
\makeatother
\begin{document}

\preprint{AIP/123-QED}

\title{Lowest order Carleman linearization  for low Reynolds long--term behaviour of fluid flow simulations}
% Force line breaks with \\
\author{Luca Cappelli~\orcidlink{0009-0009-1169-8380}}
 %\altaffiliation{Physics Department, XYZ University.}%Lines break automatically or can be forced with \\
 \email{luca.cappelli@iit.it}
\author{Sauro Succi\orcidlink{0000-0002-3070-3079}}%
\affiliation{ 
Fondazione Istituto Italiano di Tecnologia, Center for Life Nano-Neuroscience at la Sapienza, Viale Regina Elena 291, 00161 Roma, Italy
}%

\date{\today}% It is always \today, today,

\begin{abstract}
It is shown that the lowest (second) order truncation of the Carleman 
linearization of the fluid equations (C2) recovers the late stage of the 
evolution, namely the steady-state solution, although to 
a decreasing degree of accuracy at increasing Reynolds number. 
This asymptotic property is first proved analytically for the
decaying logistic with external forcing and then shown to hold to a significant
degree of accuracy also for the more complex case of two-dimensional
Kolmogorov-like fluid flow at low Reynolds numbers, below $Re \sim 10$. 
This time-asymptotic property may open interesting prospects for the quantum
simulation of low-Reynolds steady-state fluid flows.  
\end{abstract}

\maketitle

\section{Introduction}
The development of quantum algorithms for the simulation of fluid equations and transport phenomena has attracted considerable interest in recent years~\cite{SREENI, Liu2021, Succi_2026_EPL, BharadwajSreenivasan2025, EPL2023, diMolfetta2022, Tennie2025, MengZhongXu2024_SimulatingUnsteadyFlows, Schumacher24_ADR}. 
A key obstacle in this direction arises from the inherently nonlinear structure of the Navier--Stokes equations, which is incompatible with the linear framework underlying quantum simulation.
Different strategies have been proposed to circumvent this obstacle~\cite{Xue_2021, Xue2022, Tennie2024, Katz2025, LacatusMoller2026, Wang2026}; among them, Carleman linearization has emerged as the prominent approach~\cite{wiebe25, Wang2025QLBM,Costa2025, Wu2025, Zamora2026, psiq_26, Sanavio_25_PoF}, providing a systematic way of transforming a finite-dimensional nonlinear system into an infinite-dimensional linear one~\cite{Carleman}.
For all practical purposes, the infinite hierarchy is truncated (or closed)
at a finite level $K$. Increasing the truncation order generally improves
the approximation accuracy and extends the time interval over which the
truncated dynamics remains close to the exact nonlinear solution~\cite{Forets2017, Amini2025}.
Such convergence horizon decreases with increasing Reynolds number, owing to the 
growth of nonlinear interactions and small-scale activity~\cite{Frisch1995, Conde2025, Itani2022}.
Beyond such time horizon, the Carleman procedure is no longer expected 
to yield an accurate solution, and sometimes may even diverge altogether~\cite{Kowalski1991, joseph25, sanavio_CLB}.

Given the computational cost of higher--order truncations, as well as the difficulty of achieving convergence at increasing Carleman orders for complex dynamics, in this work 
we propose a change of perspective, namely to focus on the lowest--order non-trivial
Carleman approximation and investigate its long--term behaviour in 
comparison with the full nonlinear solution.
This property is first discussed in the context of a highly simplified
but representative toy model, the logistic decay equation with external forcing, and 
then demonstrated  via actual simulations of the two-dimensional Kolmogorov flow. 
This may opens up interesting prospects for the simulation of steady-state low-Reynolds
flows on quantum computers.

The paper is structured as follows.
In Sec.~\ref{sec:logistic}, we revisit the forced logistic equation and show that its long--term solution can be recovered using a second--order Carleman approximation when nonlinear effects remain sufficiently weak. This parameter plays a role analogous, though not identical, to the Reynolds number, thereby motivating the study of a forced fluid system linearized via the Carleman approach, which is introduced in Sec.~\ref{sec:carleman_of_fluids}. Finally, in Sec.~\ref{sec:numerical_results}, we present numerical simulations showing that the second--order Carleman approximation (hereafter C2) correctly recovers the long--term solution in the low Reynolds number regime. 

\section{Logistic decay with external forcing}\label{sec:logistic} 

We start by considering the logistic decay equation in the presence of an 
external forcing, namely:
\begin{equation}\label{eq:logistic_forced}
    \dot x = -ax+bx^2 + f    
\end{equation}
where $a,b>0$ and $f$ is a constant external forcing.
This dynamics admits two attractors
\begin{align}
    \label{eq:x_stable}
    &\tilde{x}_s = \frac{a}{2b} \left(1 - \sqrt{1-4g^2} \,  \right) \,;
    \\ & \label{eq:x_unstable}
    \tilde{x}_u = \frac{a}{2b} \left(1 + \sqrt{1-4g^2} \,  \right) \,,
\end{align}
where $g^2 = bf/a^2$ measures the strength of 
growth factors (nonlinearity and external force)  
versus exponential decay.
This can also be written as $(f/a)/c$ namely the ratio
of the linear long--term solution and the capacity $c=a/b$ 
of the force-free growing logistic.
The above solutions hold under the condition
$
g^2 < 1/4    
$.
\\
It is readily checked that $\tilde{x}_s$ is a stable attractor while $\tilde{x}_{u}$ is a 
repeller. Note that in the limit $g^2 \to 0$, we obtain $\tilde{x}_{s} \sim f/a$, which is 
the linear long--term solution under the effect of forcing alone.
Conversely, when $g^2 \to 1/4$ the two attractors merge at $c/2$.
and beyond such value the solution runs away.

It is of interest to compare these expressions with the long--term 
solution of the $K=2$ Carleman system.  
In order to obtain a linear system, we consider Eq.~\eqref{eq:logistic_forced} and introduce the variables $x_k \equiv x^k$. The resulting hierarchy is then truncated at order $K$, neglecting all higher--order terms. In the specific case of the $K=2$ Carleman approximation, it is readily verified that the corresponding system reads
\begin{eqnarray}\label{eq:carleman_logistic_analytic}
\dot x_1 = -ax_1+bx_2 + f \,;\\
\dot x_2 = -2ax_2+ 2fx_1 \,.
\end{eqnarray}
The logistic equation is simple enough to allow analytical solutions for the dynamics,  including the steady state solution. 
However, in more relevant cases such as the Navier--Stokes equations, the solution and its long--term behaviour are typically sought by numerical integration. 
Upon adopting a first--order explicit Euler scheme, Eqs.~\eqref{eq:carleman_logistic_analytic} take the form
\begin{eqnarray}\label{eq:carleman_logistic_CT}
x_1(t+\Delta t) = x_1 + \Delta t (f-ax_1+bx_2 ) \,;\\
x_2(t+\Delta t) = x_2 +\Delta t(2fx_1 - 2ax_2) \,,
\end{eqnarray}
where $\Delta t$ denotes the integration time step. Hereafter, we shall refer to this formulation as $CT$, indicating that the Carleman transformation is applied prior to time discretization.
At steady state $\tilde{x}_2 = f\tilde{x}_1/a$, yielding
\begin{equation}\label{eq:carleman_stable_state}
    \tilde{x}_1 = \frac{f}{a}\frac{1}{1-g^2} \,.
\end{equation}
It is nevertheless possible to derive the corresponding Carleman system by starting from the time--discretized version of Eq.~\eqref{eq:logistic_forced}. In analogy with the previous case, we shall refer to this formulation as $TC$. For convenience, and in order to facilitate comparison with the Navier--Stokes equations, we introduce the quantities $A=a\Delta t - 1$, $B=b\Delta t$, and $F=f\Delta t$. The resulting $TC$ Carleman system can then be written as
\begin{eqnarray}
    \label{eq:carleman_log_TC}
     &x_1(t+\Delta t) = -Ax_1 + Bx_2 + F \,;
    \\
     &x_2(t+\Delta t) = F^2 -2AF x_1 + (A^2 + 2BF)x_2 \,.
\end{eqnarray}
The above expression differs from the $CT$ formulation in Eq.~\eqref{eq:carleman_logistic_CT}, owing to the presence of terms of order $\mathcal{O}(\Delta t^2)$. Upon neglecting such terms, however, it is readily verified that the resulting expression becomes identical to Eq.~\eqref{eq:carleman_logistic_CT}. In other words, to first order in $\Delta t$, the operations $CT$ and $TC$ commute, and it is therefore equivalent to perform the Carleman transformation before time integration or vice versa.
Alternatively, keeping the terms $\mathcal{O}(\Delta t^2)$ leads to the following expression for the steady state solution
\begin{equation}
    \tilde{x}_1^{TC} = \frac{f}{a}\frac{1 - 0.5 \Delta t \,a(1+bg^2)}{1 - g^2 - 0.5 \Delta t \,a} \, ,
\end{equation}
which reduces to Eq.~\eqref{eq:carleman_stable_state} 
in the limit of $\Delta t \to 0$.

\begin{figure}[t]
    \centering
        \begin{subfigure}{0.45\textwidth}
        \includegraphics[width=\columnwidth]{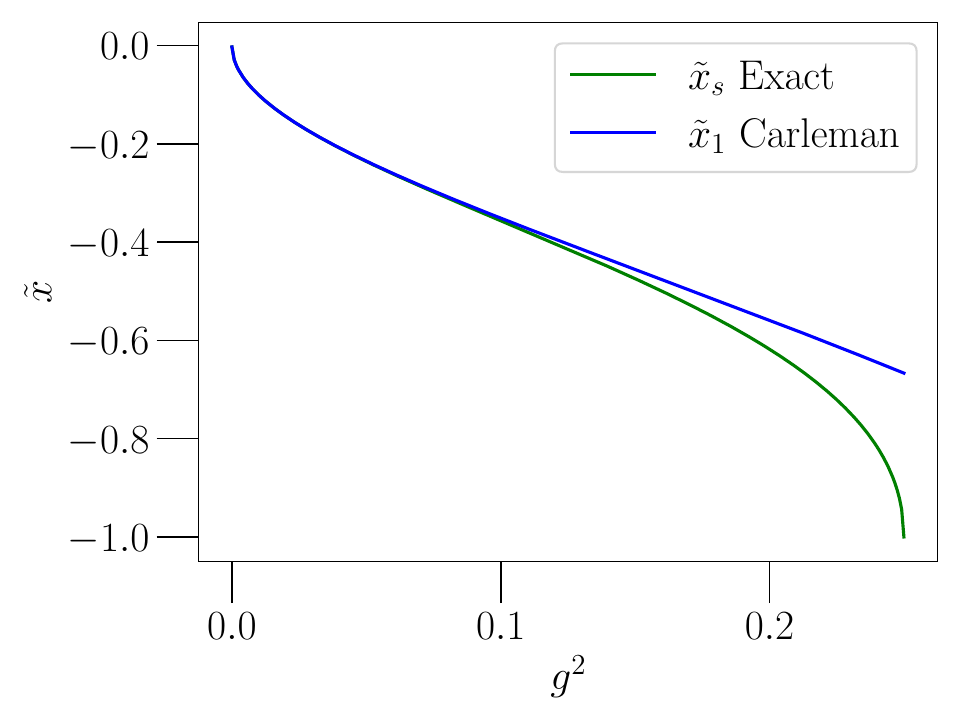}
    \end{subfigure}
    % %\raggedleft
    \begin{subfigure}{0.45\textwidth}
        \includegraphics[width=\columnwidth]{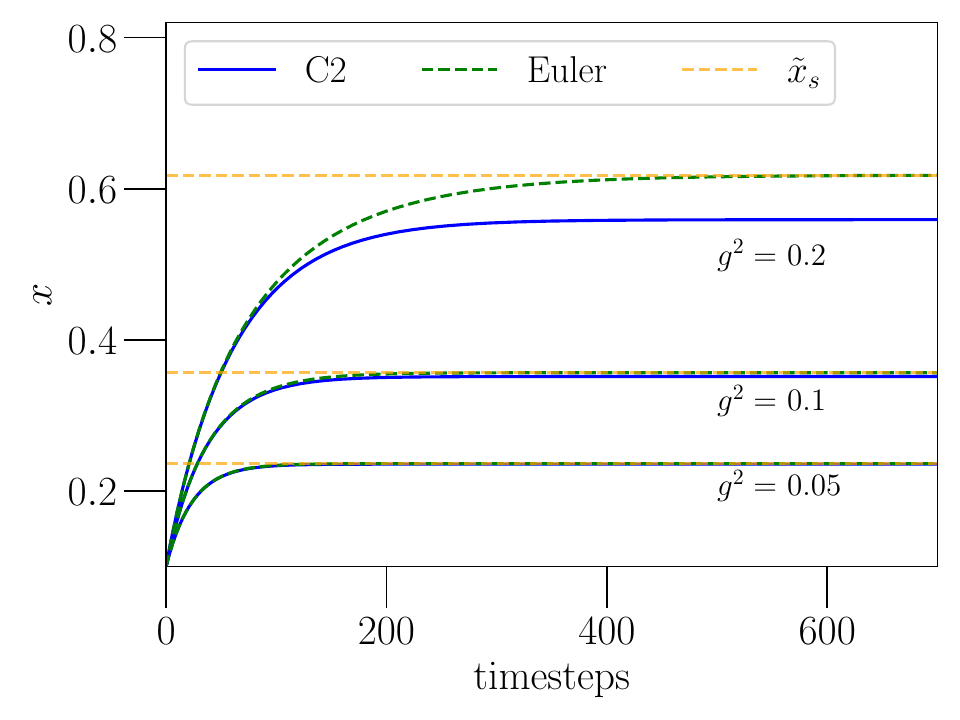}
    \end{subfigure}
    \caption{
   % \textcolor{red}{Non mostro CT e TC perche' nel testo viene spiegato che coincidono gia' dal punto di vista analitico, al primo ordine. Quello che posso fare e' mostrare come va TC con anche i termini in $dt^2$, questo giustificherebbe la scelta che abbiamo fatto poi di girare usando anche i contributi in $dt^2$.}
    (a) Stationary solutions of the forced logistic $\tilde{x}_s$ and of the 
    second--order Carleman system $\tilde{x}_1$ as a function of $g^2$. 
    Both figures show that, as $g^2$ decreases, the Carleman solution approaches both the 
    logistic and the long--term one $\tilde{x}_s$ (yellow dotted line).
    (b) Solution of the forced logistic equation~\eqref{fig:forced_logistic} with 
    Euler time-marching (green dotted line) and a second--order Carleman linearization (solid line) 
    in three different dynamical scenarios ($g^2 = 0.05,0.1,0.2$). 
    The system evolves with $\Delta t=0.01$ over $700$ timesteps, keeping $b=1, f=1$ and 
    varying $a$ accordingly to $g^2=bf/a^2$. 
    }
    \label{fig:forced_logistic}
\end{figure}

For the sake of comparison, we rewrite the 
stable attractor in Eq.~\eqref{eq:x_stable} as follows:
\begin{equation}
    \tilde{x}_s = \frac{f}{a} \frac{1-\sqrt{1-4g^2}}{2g^2} \,.
\end{equation}
Expanding to first order in $g^2$, we obtain
$\tilde{x}_s \sim f/a$, which matches the first order expansion
of Eq.~\eqref{eq:carleman_stable_state} as well. 
Hence C2 correctly reproduces the linear steady state.
Expanding to second--order in $g^2$:
$\tilde{x}_s \sim 1 + g^2 + 4 g^4$, whereas $\tilde{x}_1 \sim 1 +g^2 +g^4$, hence 
C2 incurs an error $O(g^4)$. 
This is readily seen by visual inspection of the two solutions in Fig.~\ref{fig:forced_logistic}, where for $g^2<0.1$  the steady state $\tilde{x}_s$ from Eq.~\eqref{eq:x_stable} coincides with the Carleman solution $\tilde{x}_1$ in Eq.~\eqref{eq:carleman_stable_state}.

% -------------------------------------------------------

{\it Caveat}: While the forced logistic equation bears some relevance
to the fluid equations, this analogy should not be taken too far.
In particular, the growth parameter $g^2$ should {\it not} be taken as a
literal parallel of the Reynolds number in actual fluids for 
several reasons. First, the Reynolds number does not involve
the square of the velocity (kinetic energy) but its gradient instead.
Second, actual fluid dynamics is largely affected by pressure. 
Third, dissipation is scale-selective, favouring the survival of large scales
over short ones. 
\\
Finally, and most importantly, the Navier--Stokes equations contain a pressure term, $\nabla P/\rho$, which does not vanish at steady state but plays an active role in balancing the remaining terms. 
Still in the spirit of the analogy with the logistic equation, the effect of the pressure gradient can be regarded as being incorporated into the linear operator $A$, and possibly also into $B$, thereby preventing any direct correspondence between $g^2$ and the Reynolds number. Nevertheless, it is still possible to estimate the value of $g^2$ for the Navier--Stokes equations by considering the corresponding operator norms: $g^2 = ||B||\,||F|| / ||A||^2$,where $A$, $B$, and $F$ are operators analogous to the terms appearing in the logistic equation.
\\
The logistic results provide useful hints on the behaviour of fluids but this analogy cannot be taken quantitatively.
It is therefore of interest to determine which range of Reynolds numbers corresponds to the regime in which C2 correctly reproduces the long--term solution. 
This requires a numerical analysis of the fluid equations upon Carleman linearization.

\section{Carleman linearization of the fluid equations}\label{sec:carleman_of_fluids}

In Eulerian conservative form, the Navier--Stokes equations do not lend themselves naturally to Carleman embedding~\cite{Sanavio_2024}, and in the literature alternative approaches are instead based on Carleman linearizations of lattice Boltzmann formulations~\cite{Wang2025QLBM, Sanavio_25_PoF, Zamora2026, psiq_26}.
\\
More recently, however, a formulation of the Navier--Stokes equations based on the inverse Madelung transformation~\cite{dietrich_vautherin} has shown better compatibility with Carleman linearization than the aforementioned Eulerian formulation~\cite{Cappelli26}. Moreover, in the absence of external forcing, a second--order truncation was found to recover the steady-state solution. For these reasons, we adopted this formulation to study the asymptotic behaviour in the forced regime.

Consider a two--dimensional incompressible fluid whose evolution is governed by the Navier--Stokes equations:
\begin{equation}
    \label{eq:NSE}
    \frac{\partial\mathbf{v}}{\partial t} + (\mathbf{v}\cdot\nabla)\mathbf{v}
    =
     \nu\nabla^2\mathbf{v} -\frac{\nabla P }{\rho} + \mathbf{f} \,.
\end{equation}
Here, \(\mathbf{f}(\mathbf{r})\) denotes a force field per unit mass and \(P\) is the fluid pressure. 
In particular, when the latter satisfies the ideal gas equation of state \(P = c_s^2\rho\), and 
the density can be approximated as \(\rho \approx 1\) 
under the quasi--incompressibility condition
%$|\rho-1|$ $\rho \ll 1$
, the system can 
be recast in a Hamilton--Jacobi--like form~\cite{Cappelli26, dietrich_vautherin} (NSHJ)
\begin{align}\label{eq:continuity}
    & \partial_t \rho + \nabla \cdot (\rho \, \mathbf{v}) = 0 \, ;
    \\ \label{eq:Hj}
    & \partial_t \chi + \frac{\boldsymbol{v}^2}{2} + c_s^2(\rho - 1) - (\mu + \nu) \, \nabla \cdot \boldsymbol{v} = 0
    \\
    \label{eq:evolution_of_A}
    & 
\partial_t\mathbf{A} = 
\omega
\begin{pmatrix}
v_y \\
- v_x
\end{pmatrix}
+
\nu
\begin{pmatrix}
-\partial_y \omega \\
\partial_x \omega
\end{pmatrix},
\end{align}

where the velocity field is expressed in terms of a scalar potential 
\(\chi(\mathbf{r},t)\) and a vector field \(\mathbf{A}(\mathbf{r},t)\).
The latter vector is entirely in charge of the flow   vorticity
    \label{eq:def_v}
    \begin{equation}\label{eq:definition_v}
    \mathbf{v}(\mathbf{r}, t)= \nabla \chi(\mathbf{r}, t) + \mathbf{A}(\mathbf{r}, t) \,,
\end{equation}
and $\omega=\partial_xA_y - \partial_y A_x$ represents the scalar vorticity.
Using the centered finite differences operator $D_i$ and integrating in time with an 
explicit first order Euler scheme, the system of Eqs.~\eqref{eq:continuity}~--~\eqref{eq:evolution_of_A} reads:
\begin{align}
        \label{eq:2D_continuity}
         &\hat{\rho} = \rho - \Delta t \left[
         (D_x\rho) v_x + (D_y \rho )v_y + \rho  (D_xv_x + D_yv_y) 
         \right] \, ;
         \\ \label{eq:2D_HJ}
         & \hat{\chi} = \chi + \Delta t \left[
          \nu ( D_xv_x + D_yv_y) - c_s^2(\rho - 1) - \frac{1}{2}(v_x^2 + v_y^2)
         \right] \, ;
         \\    \label{eq:2D_Ax}
         & \hat{A_x} = A_x + \Delta t \left(
                \omega  v_y - \nu D_y \omega   
         \right) \, ;
         \\    \label{eq:2D_Ay}
         & \hat{A_y} = A_y + \Delta t \left(
              \nu  D_x\omega - \omega v_x 
         \right) \,;
     \end{align}
where the hat $\hat{\cdot}$  identifies the time--evolved quantities.
This set of equations is the starting point for the Carleman embedding using the TC scheme.

Such procedure allows one to rewrite Eqs.~\eqref{eq:2D_continuity}--\eqref{eq:2D_Ay} as a linear system through the definition of the second--order lifted Carleman variables
\begin{equation}
\mathcal{J} =
\begin{pmatrix}
J^{(1)} \\
J^{(2)}
\end{pmatrix}
=
\begin{pmatrix}
J \\
J \otimes J
\end{pmatrix},
\end{equation}

where the Carleman array $J$ is given by a combination of the two dimensional fields 
\begin{equation}\label{eq:J_carleman}
    J_{\alpha} = 
    \begin{pmatrix}
        \rho \\\chi \\A_x \\A_y 
    \end{pmatrix} \,,\;\;\;\alpha=\{0,1,2,3\} \,.
\end{equation}
The second--order term $J^{(2)}_{\alpha \beta}(x,y) \equiv J_\alpha(x) \otimes J_\beta(y)$ is obtained 
through the product between the two original fields. 
With this notation, the starting system of equations takes the compact form
\begin{equation}
    \label{eq:carleman_NSHJ}
     J_\alpha(t+\Delta t) = A_{\alpha, \beta}J_\beta + B_{\alpha \beta \gamma} J^{(2)}_{\beta \gamma} + F_\alpha\,,
\end{equation}
where the constant $F_\alpha = \Delta t\,(0, c_s^2, f_x, f_y)^T $ contains the forcing terms, while the 
explicit forms of the tensors $A$ and $B$ are provided in the appendix of~\cite{Cappelli26} and are both first order terms in $\Delta t$. 
 \\
 The evolution of the second--order term follows directly from its definition and is obtained by 
 taking the tensor product of Eq.~\eqref{eq:carleman_NSHJ}:  
\begin{equation}\label{eq:second_order_carleman}
    \begin{split}
    %     \partial_tJ^{(2)}_{\alpha \beta} 
    %     & =  C\otimes C 
    % + AJ^{(1)}\otimes C + C\otimes AJ^{(1)} 
    % + (A\otimes A )J^{(2)} 
    % \\ &
    % + BJ^{(2)}\otimes C + C\otimes B J^{(2)}\,.
    J^{(2)}_{\alpha \beta}(t+\Delta t) 
        & =  F_\alpha\otimes F_\beta
    + (A_{\alpha \gamma}\otimes F_\beta + F_\alpha\otimes A_{\beta\gamma}) J_\gamma^{(1)}
    \\ &
    + (A_{\alpha \gamma}\otimes A_{\beta\delta}  +B_{\alpha \gamma \delta}\otimes F_\beta + F_\alpha\otimes B_{\beta \gamma \delta } ) J_{\gamma \delta}^{(2)}\,.
    \end{split} 
\end{equation}
The analogy with the logistic equation, Eq.~\eqref{eq:carleman_log_TC}, is now readily apparent. Upon defining $A \equiv \mathbb{I} + \Delta t\,\tilde{A}$, the above expression can be rewritten in a similar form to Eq.~\eqref{eq:carleman_logistic_CT}, retaining only terms up to first order in $\Delta t$.
\begin{align}
    \label{eq:carleman_NSHJ_first_order}
    \nonumber 
    J^{(2)}_{\alpha \beta}(t+\Delta t)  
    &= J_{\alpha\beta}^{(2)} + 
    (\mathbb{I}_{\alpha \gamma}\otimes F_\beta + F_\alpha\otimes \mathbb{I}_{\beta\gamma}) J_\gamma^{(1)}
    \\ 
    &+ 
    \Delta t(\tilde{A}_{\alpha \gamma}\otimes \mathbb{I}_{\beta\delta} + \mathbb{I}_{\alpha \gamma}\otimes \tilde{A}_{\beta\delta}) J^{(2)}_{\gamma \delta}
\end{align}

Combining Eqs.~\eqref{eq:carleman_NSHJ} and~\eqref{eq:carleman_NSHJ_first_order}, we obtain the second--order Carleman approximation of the Navier--Stokes--Hamilton--Jacobi (NSHJ) system 
of equations, corresponding to Eq.~\eqref{eq:NSE}.

%---------------------------------------------------------------------------------------------------------
\FloatBarrier
\section{Numerical simulations}\label{sec:numerical_results}
In a recent study, it has been shown that the Carleman linearization of Eqs.~\eqref{eq:2D_continuity}--~\eqref{eq:2D_Ay} recovers the long--term solution in the case of free evolution, i.e., in the absence of external forcing~\cite{Cappelli26}. While this behaviour has been previously justified in simpler cases, such as the forced logistic equation, here we aim to assess whether it persists in a more relevant setting, namely in a forced system, where the force sustains a non-zero net motion in the time asymptotic limit

\begin{figure*}[tp]
    \centering
    \includegraphics[width=.8\linewidth]{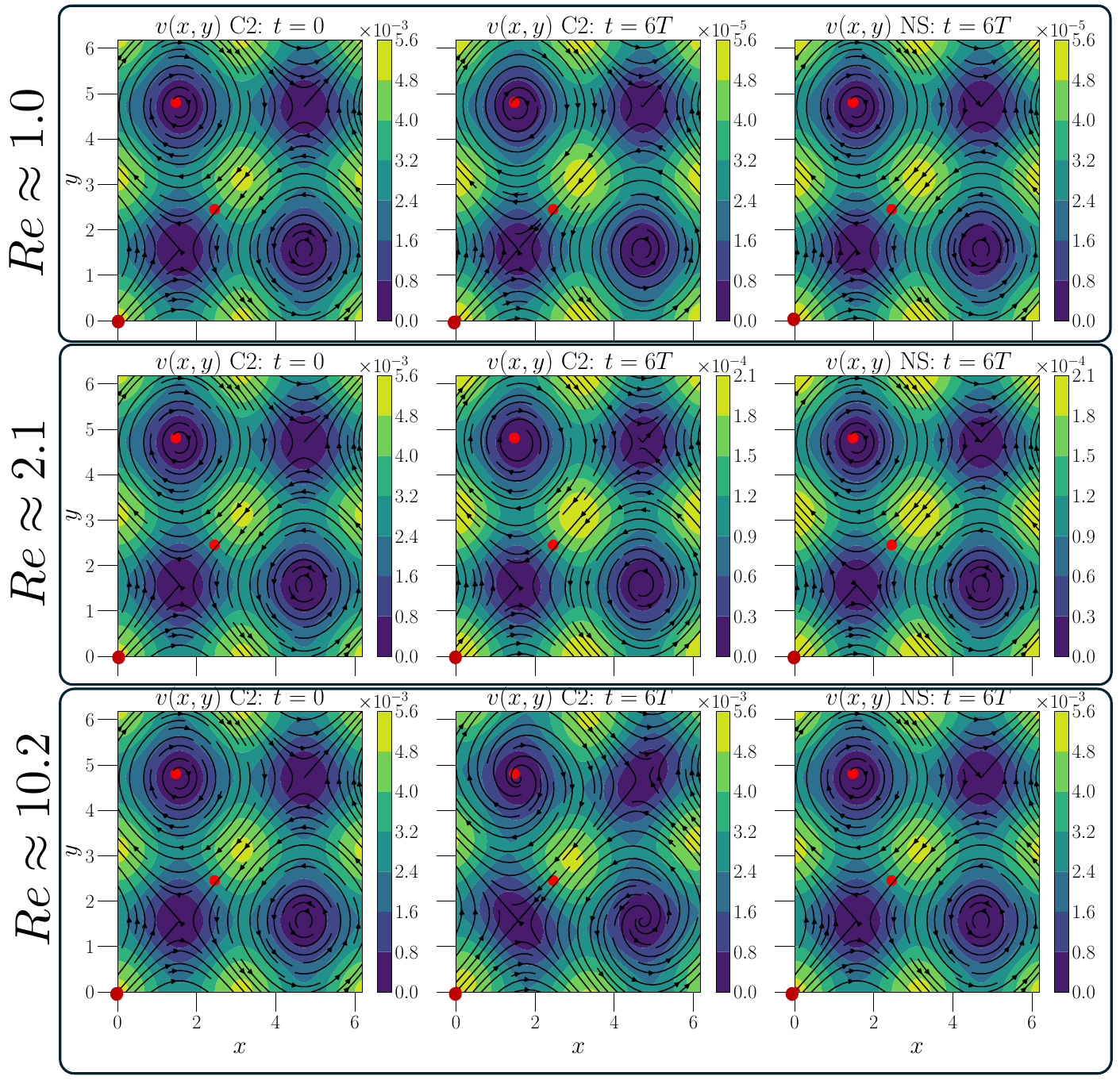}
    \caption{Amplitude and streamlines of the velocity field at $t=0$ (left-most column) and at the stationary 
    state $t=6T$ for the second--order Carleman system (center column) and  Navier-Stokes (right-most column). 
    The system refers to the initial conditions in Eq.~\eqref{eq:initial_conditions}, where each row corresponds to different Reynolds numbers. The red dots highlight the spatial locations chosen to represent the evolution of the velocity field in the figures ahead.
     % At $t=6T$, along the forced axis, C2 approximates Navier-Stokes with maximum absolute error of $\mathcal{O}(10^{-5})$ and mean $L_2$ relative error $||v_x^{_{(C_2)}} - v_x^{_{(NS)}}|| / ||v_x^{_{(NS)}}||$ of $3.2\times10^{-4}$ with $Re\approx6.6$ and $1.9\times10^{-4}$ with $Re\approx16.4$. 
    }
    \label{fig:fields}
\end{figure*}

\begin{figure}
    \centering
    \includegraphics[width=\columnwidth]{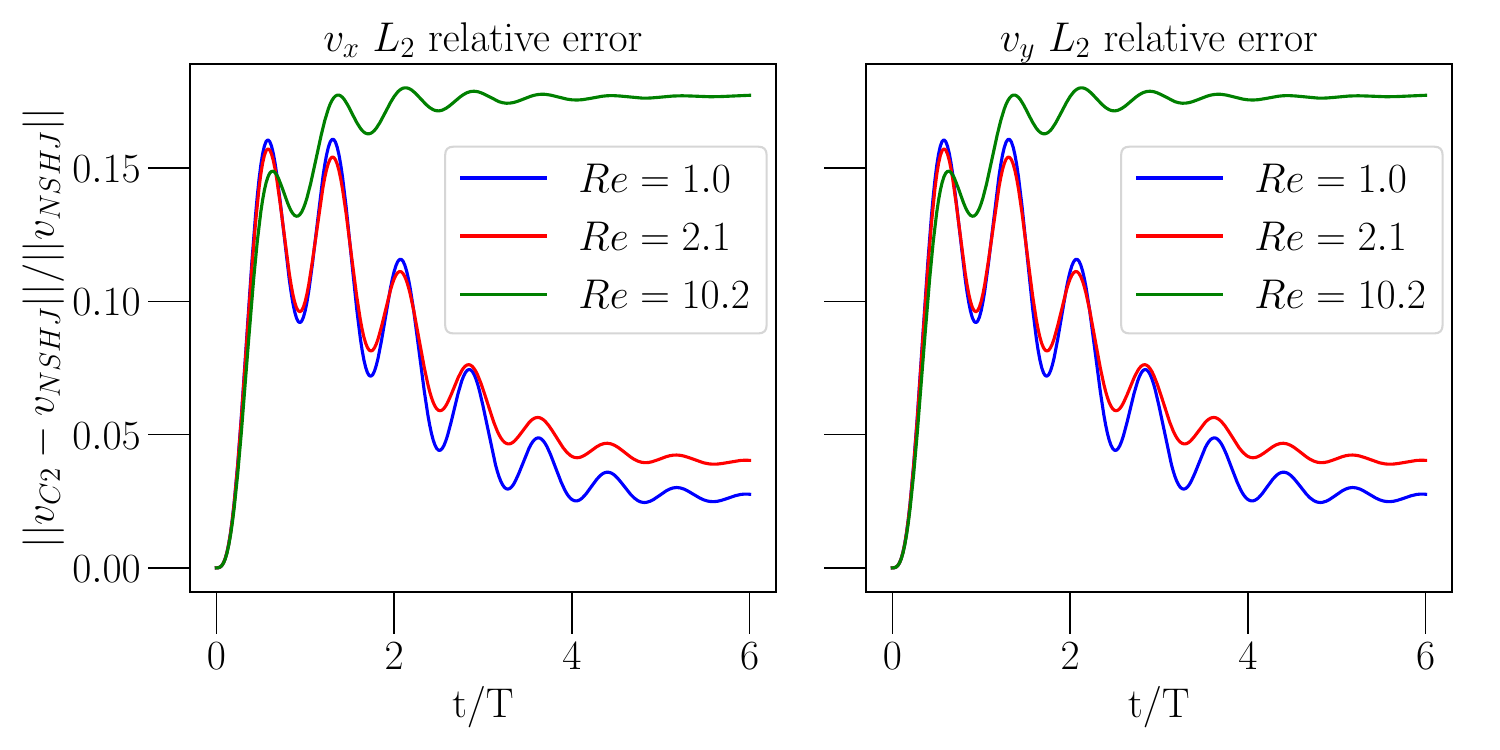}
    \caption{Global relative error of the velocity field for three different Reynolds numbers. The figure refers to the evolution of initial condition in Eq.~\eqref{eq:initial_conditions} with three different values of the forcing amplitude, respectively $f_0=5\times10^{-4}, f_0=10^{-3}$ and $f_0 = 5\times10^{-3}$. The error is evaluated using the expression in Eq.~\eqref{eq:error}. For $Re \sim \mathcal{O}(1)$ the relative error shows interesting accordance, being in the order of $5\%$ while it grows around $15-20\%$ for $Re\sim\mathcal{O}(10)$, which excludes a significant divergence from the long--term solution. For moderate Reynolds, the error saturates around $t/T \sim 4$ while the for lower Reynolds this happens around $t/T\sim 6$, suggesting that the steady state is reach towards the end of the simulation. }
    \label{fig:errors}
\end{figure}

\begin{figure*}[]
    \centering
    \includegraphics[width=.8\linewidth]{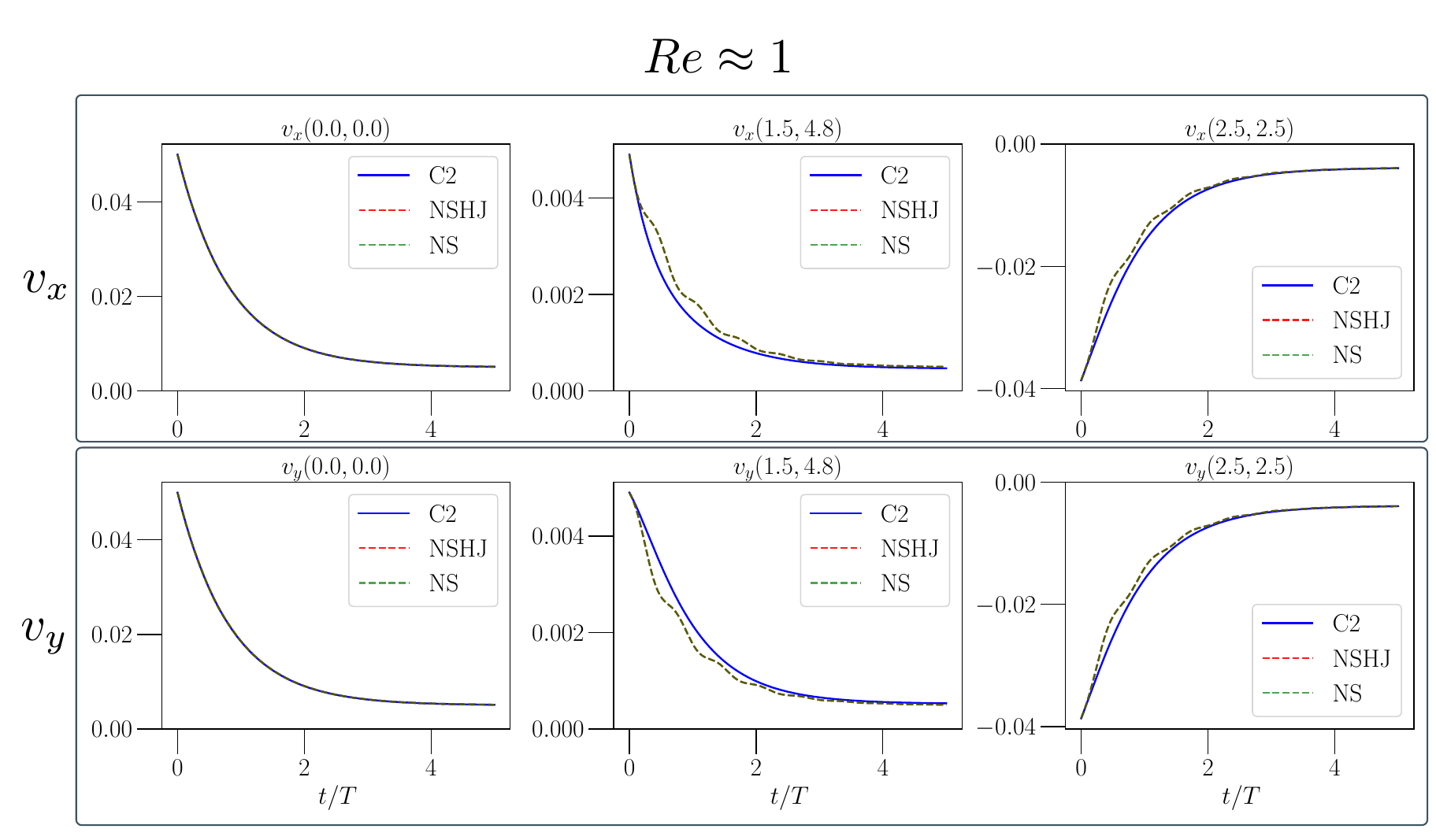}
    \caption{
    Evolution of the velocity field in three different spatial locations: $P_0 \equiv (0,0)$, where the flow attains its peak amplitude; $P_1 \equiv (1.5,4.8)$, corresponding to one of the minimum point of the velocity field and $P_2 \equiv (2.5, 2.5)$, where the flow attains negative values. These points correspond to the red dots shown in Fig.~\ref{fig:fields}. The initial condition in Eq.~\eqref{eq:initial_conditions}, with $f_0 = 5\times10^{-4}$,  is evolved on a $64\times64$ grid for $3\,000$ timesteps with $\Delta t = 0.02$. The C2 scheme recovers the long--term solution perfectly for each chosen point. The two reference solutions, Navier--Stokes and NSHJ (dashed lines) are equal, overlapping each other.}
    \label{fig:velocities_re_1}
\end{figure*}

\begin{figure*}[]
    \centering
    \includegraphics[width=.8\linewidth]{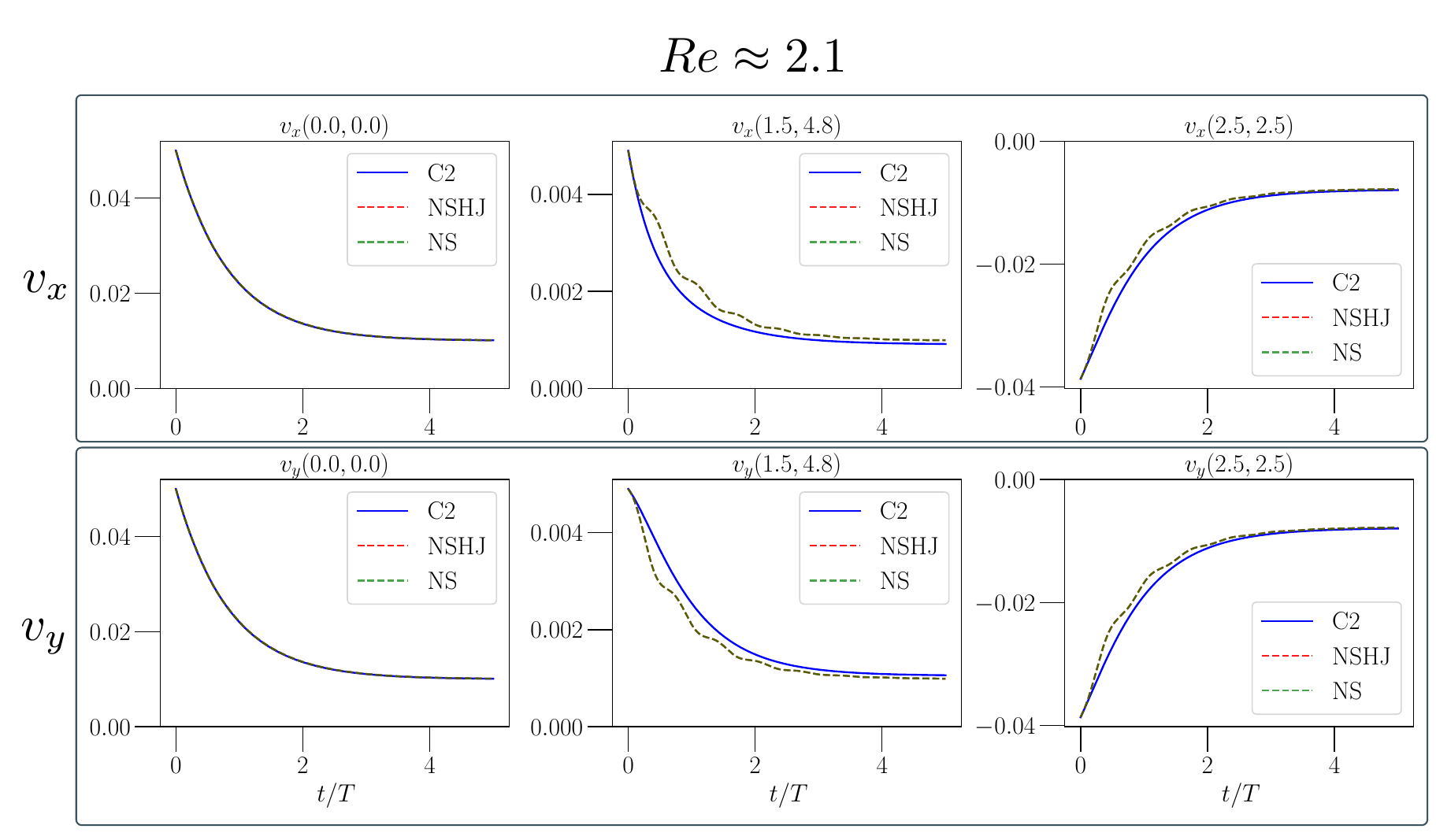}
    \caption{
    Evolution of the velocity field in three spatial locations of the previous figures. 
    The initial condition in Eq.~\eqref{eq:initial_conditions} is evolved with $f_0 = 10^{-3}$ on a $64\times64$ grid over $3\,000$ timesteps with $\Delta t = 0.02$. 
    While the C2 solution recovers the correct long--term solution, minor discrepancies appears for $P_1=(1.5, 4.8)$. Even in this scenario the Navier--Stokes solution and the NSHJ lines overlaps perfectly.}
    \label{fig:velocities_re_2}
\end{figure*}

\begin{figure*}
    \centering
    \includegraphics[width=.8\linewidth]{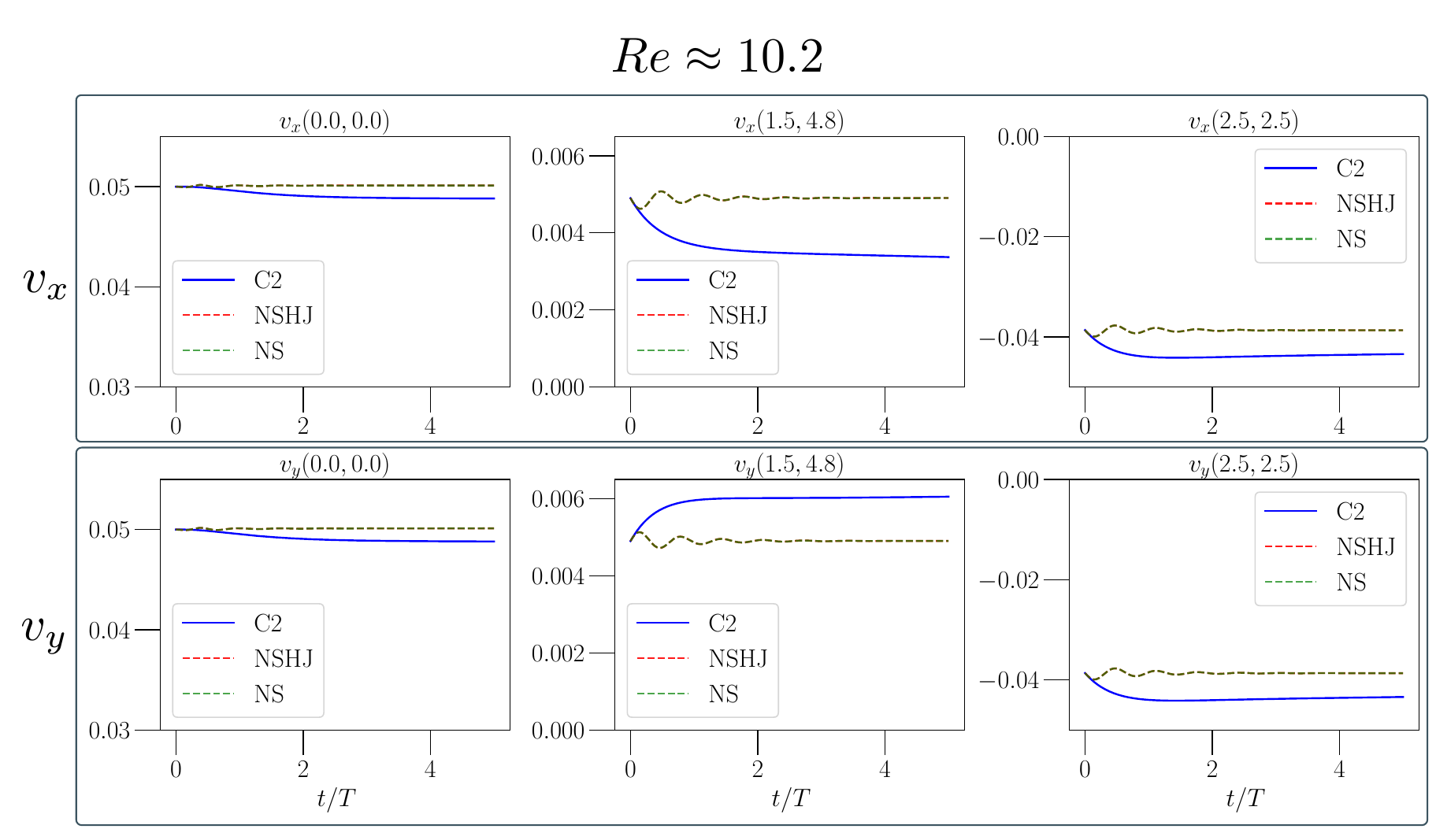}
    \caption{
    Evolution of the velocity field along the $x$--axis (top row) and the $y$--axis (bottom row). 
    The figure refers to the initial condition in Eq.~\eqref{eq:initial_conditions} evolved with $f_0 = 5\times10^{-3}$ for $3\,000$ timesteps, with $\Delta t=0.02$ on a $64\times 64$ grid. The spatial probes correspond to the red dots in Fig.~\ref{fig:fields}. 
    In this scenario, where $Re\approx10.2$, C2 does not recover the long--term solution, however it does not diverge completely, showing a relative error of approximately $2\%$ for $P_0=(0,0)$, $20\%$ for $P_1(1.5, 4.8)$ and $15\%$ for $P_2=(2.5,2.5)$. These numbers are in agreement with the global error shown in Fig.~\ref{fig:errors}.
    }
    \label{fig:velocities_re_10}
\end{figure*}

We consider a fluid system defined on a periodic square grid with \(G = N \times N = 64^2\) 
points and box size \(L_{\text{box}} = 2\pi\). 
The initial condition is chosen to represent a Kolmogorov-like flow:
\begin{align}\label{eq:initial_conditions}
    \rho = 1, \quad \chi = 0, \quad
    A_x = u_0 \cos(y), \quad A_y = u_0 \cos(x),
\end{align}
with $u_0=0.05$ and the forcing acting along one component, analogously to the initial velocity, namely \(f_x = f_0 \cos(y)\) and \(f_y = f_0 \cos (x)\). 
By varying the amplitude of the forcing terms, we aim to study the same initial condition under different Reynolds numbers. We also fix the speed of sound to \(c_s^2 = 1/3\) and the viscosity to \(\nu = 0.1\). 
\\
All the cases studied in this work reach a stationary velocity value $U_s$ through the stabilization of the forcing term;
thus the Reynolds number is defined accordingly, in terms of the stationary velocity rather than the initial amplitude:
\begin{equation}\label{eq:reynolds}
    Re = \frac{(U_{s} / \pi) N}{\nu} \,,
\end{equation}
where \(2U_s / \pi\) denotes its mean value, expressed in dimensionless code units and the size is $N/2$ since we consider the wavenumber $k=1$.

In this section, we investigate the performance of the second--order Carleman approximation at low and moderate Reynolds numbers by varying the amplitude of the forcing term. In particular, we consider three values of the forcing amplitude. Starting from $f_0=5\times10^{-4}$, corresponding to a Reynolds number of approximately $Re\simeq1.0$, we then double the forcing amplitude, obtaining $Re\simeq2.1$. Finally, with $f_0=5\times10^{-3}$, we enter an intermediate Reynolds-number regime characterized by $Re\simeq10.2$.
\\
For all cases, we fix $T/\Delta t = 500$ and evolve the system for more than $3\,000$ timesteps. Here, $T=1/(\nu k^2)$ denotes the diffusive timescale of the flow, corresponding 
to a timestep $\Delta t=0.02$.

% ---------------------        STREAMLINES
Figure~\ref{fig:fields} shows both the velocity magnitude and the corresponding streamlines at $t=0$ and after six characteristic times, $t=6T$, for the three cases considered above. The figure provides a qualitative comparison between the long--term solutions obtained with the second--order Carleman approximation and those computed from the full Navier--Stokes equations.
\\
Two distinct behaviours can be identified. For Reynolds numbers of order unity, excellent visual agreement is observed between the Carleman solution and the reference Navier--Stokes state. Both the streamline topology and the velocity magnitude are reproduced with remarkable accuracy. In contrast, for the case $Re\simeq10.2$, noticeable discrepancies begin to emerge. The spatial distribution of $|\mathbf{v}|^2$ no longer matches the reference solution, exhibiting elongated structures in the bottom--left and top--right corners of the domain. Moreover, small spurious vortical structures appear in regions where the velocity is weakest. Despite these differences, it is worth emphasizing that the overall flow pattern remains qualitatively similar to the long--term reference solution, with no evidence of an abrupt departure from the Navier--Stokes dynamics.

%  --------------------         ERRORS
A more quantitative assessment of the results is provided in Fig.~\ref{fig:errors}, which reports the evolution of the global relative error for the velocity fields along the two spatial directions.
\\
The error is computed using the \(L_2\) norm and normalized by the mean magnitude of the field rather than point-wise, in order to avoid artificially large relative errors in regions where the velocity approaches zero. The corresponding expression reads
\begin{equation}
    \label{eq:error}
    \frac{||\Delta v_i||_{L_2}}{||v_i||_{L_2}}
    =
    \frac{|| v_i^{C2} - v_i^{NS} ||_{L_2}}
         {|| v_i^{NS} ||_{L_2}}
    \,,
\end{equation}
where $i=\{x,y\}$. The superscript C2 denotes the solution obtained through the second--order Carleman approximation, while NS refers to the reference Navier--Stokes solution.
\\
Once again, it is observed that the good visual agreement obtained at Reynolds numbers of order unity translates into global errors of approximately $5\%$. Considering the small magnitude of the velocity field, especially in the case $Re \approx 1$, this result further confirms what was observed for the logistic equation, namely that C2 is capable of recovering the long--term solution with good accuracy.
As the forcing amplitude is increased, however, the error grows and eventually reaches a global value of approximately $16\%$. This finding is consistent with the visual comparison shown in Fig.~\ref{fig:fields} and confirms that, as the Reynolds number increases, C2 no longer reproduces the exact solution but instead incurs an error that scales with $g^2$. Once again, the analogy with the logistic equation is instructive, although $g^2$ should not be interpreted as the literal Reynolds number.
\\

% %--------------         RE=1
Finally, we examine the temporal evolution of the velocity field at selected spatial locations in order to gain a more detailed understanding of the dynamics. To this end, we consider the points highlighted in red in Fig.~\ref{fig:fields}. These correspond respectively to the locations of maximum and minimum velocity, $P_0 \equiv (0,0)$ and $P_1 \equiv (1.5,4.8)$, the latter coinciding with the region where spurious vortices emerge for $Re \approx 10.2$. We also consider $P_2 \equiv (2.5,2.5)$, a representative point where the velocity attains negative values. Only probes located in the left half of the domain are shown, since the flow is symmetric under reflection about the axes \(x=\pi\) and \(y=\pi\).

Figure~\ref{fig:velocities_re_1} reports the evolution of the velocity field at these three representative locations for $f_0=5\times10^{-4}$, corresponding to $Re\approx1.0$. The first row shows the evolution of $v_x$, while the second row reports $v_y$. The symmetry of the system is preserved and the Carleman approximation accurately reproduces the long--term solution. As expected, however, the transient oscillatory behaviour is not captured.

A similar behaviour is observed in Fig.~\ref{fig:velocities_re_2}, corresponding to $f_0=10^{-3}$ and $Re\approx2.1$. The long--term solution is again correctly recovered, while the transient dynamics remain absent. In this case, however, a small but noticeable offset develops at point $P_1$, where the velocity reaches its minimum value. This observation is consistent with the larger global relative error reported in Fig.~\ref{fig:errors}.

Finally, Fig.~\ref{fig:velocities_re_10} shows the evolution of the flow for $f_0=5\times10^{-3}$, corresponding to $Re\approx10.2$. This case may be regarded as representative of a moderate Reynolds-number regime. Here, the long--term solution is no longer recovered exactly, although the Carleman approximation still exhibits a relatively modest departure from the reference solution. While the Navier--Stokes dynamics undergo damped oscillations before stabilizing around the initial velocity amplitude, the Carleman solution gradually drifts away from the steady state: noticeably in $P_1$ (center column) $v_x$ tends to decay, whereas $v_y$ exhibits a gradual increase. This is symptomatic of the spurious vortex we observed in the Carleman solution in Fig.~\ref{fig:fields}.
\\
Nevertheless, the error remains bounded and eventually reaches a stationary value, in a manner qualitatively similar to the logistic case with $g^2=0.2$ shown in Fig.~\ref{fig:forced_logistic}. 
\\
Again, $g^2$ is to be intended as a Reynolds-like quantity and not $Re$ itself. Nevertheless, these numerical test confirm what has been observed with the logistic equation: C2 is able to recover the steady state with an error $\mathcal{O}(g^4)$, which in this case corresponds to low Reynolds number close to unity. 

\section{Summary}

The Carleman approximation allows a nonlinear system to be cast into a linear form, provided that a truncation (or closure) is introduced at order $K$ in the lifted variables. Increasing $K$ generally improves accuracy and extends the temporal horizon over which the approximation remains valid.
\\
Here, we adopt a different viewpoint. Instead of increasing the truncation order, we restrict ourselves to the lowest nontrivial closure, $K=2$, and investigate its ability to capture the long--term dynamics of the system.
\\
Building on analytical insights from the decay logistic equation with external forcing, we have shown that a second--order Carleman approximation recovers the long--term solution of fluid flows at low Reynolds numbers.

The resulting error emerges as a correction proportional to the square of the
Reynolds-like quantity $g^2$.
We then tested these ideas on the physically relevant case of a two-dimensional 
fluid flow under periodic forcing. 
Using the second--order Carleman approximation of the Navier--Stokes--Hamilton--Jacobi equations~\cite{Cappelli26}, we studied the asymptotic behaviour of a two-dimensional forced Kolmogorov-like flow, for various Reynolds numbers up to $Re \sim 10$.
Across all the examples considered in this work, the second--order Carleman approximation accurately captures the long--term behaviour of the 2D incompressible, viscous fluid moving under the effect of the external field only in for low Reynolds numbers.
%---sistema da qui in poi
For moderate Reynolds numbers, we observe that the Carleman solution approximates the exact solution with a nearly constant relative error of about $16\%$, consistently with 
the hypothesis that C2 recovers the long--term dynamics with $O(g^4)$ accuracy.
\\
In \cite{Liu2021}, the authors report similar theoretical limitations, however with the Burgers equation they are able to reach regimes up to $Re\approx 40$ using a fourth--order Carleman approximation. In the Navier--Stokes equations, however, the pressure term limits the validity of the analogy with $g^2$ and ultimately also with the Burgers equation.

These results suggest a potentially significant shift in perspective: the second--order Carleman approximation may be interpreted as a tool for computing long--time fluid dynamics within a false--transient framework. This stands in contrast with the standard paradigm, which is primarily concerned with extending the time-horizon of the short-term dynamics. 

At present, the observed limitation in the accessible Reynolds number is mainly 
a consequence of memory restrictions on classical computers, but the loss
of accuracy at increasing Reynolds number, although still comparatively
modest at $Re \sim 10$, is nonetheless appreciable.  

Future developments will therefore focus on numerical investigations based on a Carleman--lattice Boltzmann formulation, where the accuracy limitations are related to the 
Mach rather than the Reynolds number, at least for Reynolds numbers below $100$.

Finally, we wish to observe that the present range of Reynolds numbers
is far below turbulent regimes but nonetheless relevant to other complex flow 
applications, such as flows in porous media and biological flows~\cite{DULLIEN19751, Cali92, Montessori_2019}.

\section{Acknowledgments}
The authors acknowledge valuable discussions with M. Durve, A. Jonnalagadda, M. Lacatus, M.Bernaschi, A. Roggero and A. Zecchi. 
The authors have no conflicts to disclose. 
The data that support the findings of this study are available from the corresponding 
author upon reasonable request.

\FloatBarrier

\nocite{*}
\bibliography{biblio}% Produces the bibliography via BibTeX.

\end{document}